\RequirePackage{fix-cm}
\documentclass[aps,pra,reprint,superscriptaddress,longbibliography,nofootinbib,floatfix]{revtex4-2}
\usepackage{amsmath,amssymb,mathrsfs,bm,graphicx,booktabs}
\usepackage{placeins}
\usepackage[breaklinks=true,hidelinks]{hyperref}
\newcommand{\cL}{\mathcal L}
\newcommand{\cD}{\mathcal D}
\newcommand{\Tr}{\operatorname{Tr}}
\newcommand{\ket}[1]{|#1\rangle}
\newcommand{\bra}[1]{\langle #1|}
\newcommand{\ord}{\mathcal O}
\newcommand{\PaperTitle}{Diagnostic-dependent lasting Mpemba effect in a parametrically driven oscillator}
\graphicspath{{figures/}}
\hypersetup{pdftitle={Diagnostic-dependent lasting Mpemba effect in a parametrically driven oscillator},pdfauthor={Shu He}}

\begin{document}
\raggedbottom
\title{\PaperTitle}
\author{Shu He}
\email{heshu@sicnu.edu.cn}
\affiliation{Department of Physics and Electronic Engineering, Sichuan Normal University, Chengdu 610066, China}
\date{October 27, 2026}
\begin{abstract}
For two Gaussian preparations of a detuned degenerate parametric amplifier below threshold, the trace distance shows a lasting Mpemba reversal, whereas the relative entropy eventually restores the initial order and the Hilbert--Schmidt distance alternates indefinitely. All three diagnostics agree on the initial order, and these orders persist on open sets of preparations and quadratic generators. The disagreement is possible because a real relaxation mode and a complex-conjugate pair share the leading decay rate, a coincidence that holds for any single-mode quadratic oscillator with linear damping in the elliptic regime. Number symmetry hides their competition in the undriven trace-distance tail, and parametric driving exposes it. Near a pure stationary state, the relative entropy strongly weights changes of the symplectic eigenvalue, which approximately follow the nonoscillating energy deviation; thermal occupation weakens this preference. A first-order splitting rule shows that weak number dephasing and two-photon loss favor the real branch, whereas a Kerr interaction with the same sign as the detuning favors the pair. We quantify the thermal-occupation and dephasing tolerances and estimate the measurement precision needed to observe the disagreement. Whether a Mpemba reversal is lasting therefore depends on the diagnostic and on the observation window.
\end{abstract}
\maketitle
\section{Introduction}\label{sec:intro}
The Mpemba effect occurs when a preparation initially farther from stationarity overtakes a closer one during relaxation~\cite{Lu2017,Klich2019,Carollo2021,Ares2025Review}. Trapped-ion experiments have demonstrated inverse, symmetry-restoring, strong, and multiple-crossing forms of this effect~\cite{Shapira2024,Joshi2024,Zhang2025,Xia2026Multi}. These developments raise an operational question: does observing a system for longer settle which preparation is closer to its stationary state?

When one simple real eigenvalue sets the slowest decay, the preparation amplitudes fix a common late-time order for normalized operator-convex Petz divergences under the conditions established by Salazar~\cite{Salazar2026}. The parametrically driven oscillator studied here has a different leading spectrum. Its parity-even sector contains one real eigenoperator and a complex-conjugate pair with exactly equal decay rates, so their relative phase keeps rotating while all three contributions remain comparable. This coincidence is structural: it holds for every single-mode quadratic Hamiltonian with linear damping in the stable elliptic regime.

Two lines of work bear on this question: Mpemba effects in oscillators and repeated crossings. Longhi studied Fock, squeezed, and cat preparations~\cite{Longhi2024Bosonic} and moment conditions for accelerated oscillator thermalization~\cite{Longhi2025}. Symmetry organizes Liouvillian sectors and constrains the relaxation modes accessible to a preparation~\cite{Buca2012,Albert2014,Summer2026}. In a driven-dissipative qubit, Chatterjee \emph{et al.} find that exceptional points and complex Liouvillian eigenvalues can produce multiple crossings of observables, while the Kullback--Leibler divergence from stationarity shows only one crossing~\cite{Chatterjee2024}. Ma and Liu use long-time mode analysis to distinguish persistent crossings from lasting reversals in a parity-time-symmetric qubit~\cite{MaLiu2025}. Related studies address parametrically driven coupled oscillators~\cite{Pahlevani2026}, collective reordering in degenerate bosonic modes~\cite{Araujo2026Bosonic}, classical underdamped relaxation~\cite{Shapira2026Underdamped}, and initial system--bath slip~\cite{Crotti2026}. Dos Santos \emph{et al.} analyze a resonant parametric oscillator whose quadratures decay at distinct real rates~\cite{Santos2026RLC}; in the elliptic detuned regime their real parts coincide. Chirico \emph{et al.} find that the weak-coupling spin--boson model can show a Mpemba effect in trace distance but no crossing in quantum relative entropy at low temperature~\cite{Chirico2026}. These results motivate asking how competing modes determine the eventual order when different diagnostics agree on which preparation starts farther away.

Here we show that a single pair of Gaussian preparations can have three different eventual orders under the same dynamics. The trace distance gives a lasting Mpemba reversal, the relative entropy eventually restores the initial order, and the Hilbert--Schmidt (HS) distance keeps alternating, although all three agree initially [Fig.~\ref{fig:one}(a)]. These orders persist on open sets of preparations and quadratic generators. A real mode carries the nonoscillating energy deviation, while a conjugate pair carries rotating squeezing deviations. Their common decay rate allows each diagnostic's weighting of these contributions to determine its eventual order. We explain the enhanced radial weight of relative entropy near a pure stationary state, derive a first-order rule for selection by weak perturbations, quantify the thermal and dephasing tolerances, and estimate the measurement precision. Weak number dephasing lifts the rate coincidence and eventually selects the real mode, restoring a common order; we determine the intervening observation window. The model is a detuned degenerate parametric amplifier, with direct optical and superconducting-circuit realizations.

\section{Model, preparations, and diagnostics}\label{sec:model}
With $\hbar=1$, the quadratic generator is
\begin{align}
 H_\xi&=\omega n+\xi(a^2+a^{\dagger2}),\qquad n=a^\dagger a,\label{eq:H}\\
 \cL_0\rho&=-i[H_\xi,\rho]+\kappa\cD[a]\rho,\label{eq:L0}\\
 \cD[L]\rho&=L\rho L^\dagger-\tfrac12\{L^\dagger L,\rho\}.
\end{align}
These equations describe a detuned degenerate parametric amplifier below threshold, in the frame rotating at half the pump frequency~\cite{Collett1984,Beaulieu2025}. We take $\kappa>0$ and work in the elliptic regime $2|\xi|<|\omega|$, where the Hamiltonian phase-space orbits are ellipses, with $\nu=\sqrt{\omega^2-4\xi^2}$. The preparations satisfy $\Pi\rho\Pi=\rho$, where $\Pi=e^{i\pi n}$. This operator-parity condition excludes coherences between even and odd photon numbers but allows population in both parity sectors; the dynamics preserves it.

The two signs of the detuning are related by the antilinear map $\rho\mapsto U\rho^*U^\dagger$, where $U=e^{i\pi n/2}$ and the conjugation is in the Fock basis. Since $Ua^2U^\dagger=-a^2$, this map sends $(\omega,\xi,K)$ to $(-\omega,\xi,-K)$ when the Kerr term $(K/2)n(n-1)$ is included. It leaves all dissipative channels considered below invariant, complex-conjugates the Liouvillian spectrum, and preserves the three diagnostics. The preparation phase transforms as $\varphi\to\pi-\varphi$. We take $\omega>0$ and quote rates in units of $\omega$.

For $a=(Q+iP)/\sqrt2$ and $[Q,P]=i$, preparation $i\in\{A,B\}$ is the centered Gaussian state $G[V_i]$ with
\begin{align}
 V_i&=\begin{pmatrix}n_i+\frac12+\operatorname{Re} m_i&\operatorname{Im} m_i\\\operatorname{Im} m_i&n_i+\frac12-\operatorname{Re} m_i\end{pmatrix},\label{eq:V}\\
 m_i&=\frac{n_i}{\sqrt2}e^{i\varphi_i}.\nonumber
\end{align}
These squeezed thermal states have $\langle n\rangle=n_i$, $\langle a^2\rangle=m_i$, and symplectic eigenvalue $\sqrt{n_i^2/2+n_i+1/4}$; their principal axes are rotated by $\varphi_i/2$. Our main parameters are
\begin{equation}
\begin{gathered}
 \xi=0.1\omega,\quad\kappa=0.5\omega,\quad
 (n_A,\varphi_A)=(0.5,\pi),\quad\varphi_B=0.
\end{gathered}\label{eq:family}
\end{equation}
We scan $n_B$ and call $n_B=0.33$ the principal pair. Conventions and explicit covariances are collected in Appendix~\ref{app:gaussian}.

We compare three diagnostics of the distance to the stationary state $\rho_{\rm ss}$ of the generator under consideration: the trace distance $D_i=\frac12\|\rho_i-\rho_{\rm ss}\|_1$, the HS distance $D_{{\rm HS},i}=\|\rho_i-\rho_{\rm ss}\|_2$, and the Umegaki relative entropy $\mathscr S_i=\Tr\rho_i(\log\rho_i-\log\rho_{\rm ss})$. The first and third are contractive under quantum channels. The HS distance provides a simple noncontractive comparison whose square is quadratic in the density-operator deviation. For each diagnostic $X$ we write $R_X(t)=X_A(t)/X_B(t)$, for example $R_{\rm tr}=D_A/D_B$.

A \emph{lasting reversal} requires $X_A(0)>X_B(0)$ and $X_A(t)<X_B(t)$ at all sufficiently late times~\cite{Caldas2026}. Moroder \emph{et al.} call the effect genuine when the preparation with the higher nonequilibrium free energy, equivalently the larger relative entropy to the Gibbs state, relaxes faster~\cite{Moroder2024}; it has also been observed experimentally~\cite{Schnepper2026}. Here the stationary state is driven, so we apply the lasting-reversal criterion to each diagnostic separately. The ordering is \emph{persistently alternating} if $X_A-X_B$ changes sign at arbitrarily late times, and \emph{eventually nonreversed} if $X_A>X_B$ at all sufficiently late times. We say that the late-time order places A closer, alternates, or places A farther; when A starts farther, these are the three orders just defined. Each diagnostic is judged against its own initial order. The principal initial trace distances are $0.3011$ and $0.2831$; they cross at $\omega t=1.36$, where both equal $0.182$. Numerical methods and convergence checks are in Appendix~\ref{app:numerics}.

\section{Symmetry and the periodic trace-distance tail}\label{sec:tail}
Second moments give a closed description of the Gaussian relaxation. Let $J=\left(\begin{smallmatrix}0&1\\-1&0\end{smallmatrix}\right)$ and $\mathsf F_H=J\operatorname{diag}(\omega+2\xi,\omega-2\xi)$. The drift and symplectic evolution are
\begin{equation}
 \mathsf F=\mathsf F_H-\tfrac\kappa2I,\qquad S(t)=\cos(\nu t)I+\frac{\sin(\nu t)}\nu \mathsf F_H.\label{eq:S}
\end{equation}
The stationary covariance $W$ solves $\mathsf F W+W\mathsf F^T+\kappa I/2=0$. With $\Delta_i(t)=S(t)(V_i-W)S^T(t)$, exact Gaussian evolution gives $V_i(t)=W+e^{-\kappa t}\Delta_i(t)$~\cite{Weedbrook2012,HolevoWerner2001}.

\begin{figure*}[t]
\includegraphics[width=\textwidth]{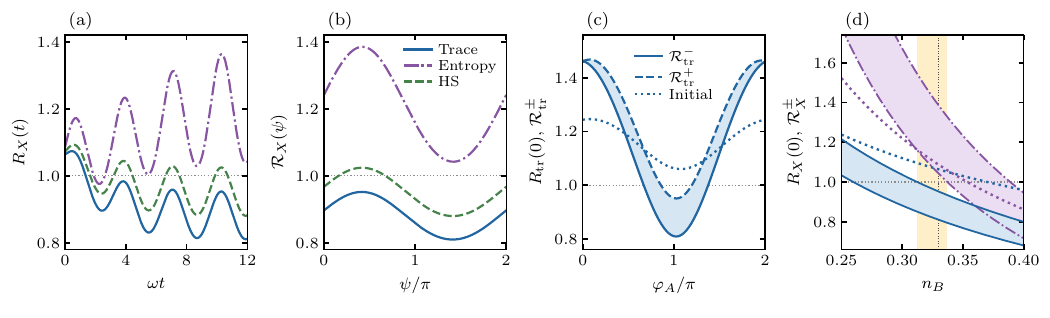}
\caption{Eventual orders for the main parameters, Eq.~\eqref{eq:family}.
(a) Finite-time ratios at $n_B=0.33$: trace distance (blue solid), relative entropy (purple dash-dotted), and HS distance (green dashed). After the entropy return at $\omega t=2.62$, trace distance and entropy keep opposite orders, while the HS ratio keeps crossing unity.
(b) Corresponding leading periodic ratios versus $\psi/\pi$, where $\psi=2\nu t$.
(c) Trace-ratio extrema $\mathcal R_{\rm tr}^-$ (solid) and $\mathcal R_{\rm tr}^+$ (dashed), and the initial ratio (dotted), versus the preparation phase $\varphi_A$ at $n_B=0.33$. Horizontal dotted lines mark equal distances.
(d) Leading intervals $[\mathcal R_X^-,\mathcal R_X^+]$ of the trace distance (blue) and relative entropy (purple) versus $n_B$, with the corresponding initial ratios (dotted). The pale vertical band marks opposite lasting orders, $0.3124<n_B<0.3371$; the vertical dotted line marks $n_B=0.33$.}\label{fig:one}
\end{figure*}

For centered Gaussian inputs and parity-invariant inputs of finite Fock support, the trace-norm expansion (Appendix~\ref{app:tail}), with a remainder uniform in the oscillation phase, is
\begin{align}
 \rho_i(t)-\rho_{\rm ss}&=e^{-\kappa t}T_i(t)+\ord(e^{-2\kappa t}),\label{eq:tail}\\
 T_i(t)&=-\tfrac12\sum_{ab}(J\Delta_i(t)J^T)_{ab}[r_a,[r_b,\rho_{\rm ss}]],\label{eq:T}
\end{align}
where $\bm r=(Q,P)^T$. The tail evolves in the three-dimensional space of second-moment deviations, where the generator has eigenvalues
\begin{equation}
 -\kappa,\qquad-\kappa\pm2i\nu.\label{eq:triple}
\end{equation}
These are pairwise sums of the first-moment eigenvalues $-\kappa/2\pm i\nu$. The equality of their real parts is structural. For any single-mode quadratic Hamiltonian with linear Lindblad operators $L_j=u_ja+v_ja^\dagger$, the drift is $\mathsf F_H-\kappa_{\rm eff}I/2$ with $\kappa_{\rm eff}=\sum_j(|u_j|^2-|v_j|^2)>0$ for stability, and in the elliptic regime the second-moment eigenvalues are $-\kappa_{\rm eff}$ and $-\kappa_{\rm eff}\pm2i\nu$~\cite{Prosen2010,Barthel2022}; diffusion changes $W$ and the norm weights but not these rates (Appendix~\ref{app:tail}). The three contributions therefore remain comparable at late times, and symmetry determines whether their competition changes the trace norm.

At $\xi=0$, number rotation commutes with attenuation and the stationary state is the vacuum. The leading operator rotates unitarily, leaving its trace norm independent of phase:
\begin{equation}
 D_i(t)=e^{-\kappa t}C_i+\ord(e^{-2\kappa t}),\quad
 C_i=\frac{n_i+\sqrt{n_i^2+2|m_i|^2}}2.\label{eq:protected}
\end{equation}
The late-time ratio is the constant $C_A/C_B$. An initial order $D_A(0)>D_B(0)$ with $C_A<C_B$ then produces a lasting Mpemba reversal. Lasting reversals have also been found for the undriven oscillator with other preparations~\cite{Longhi2024Bosonic,Longhi2025}. Symmetry fixes the tail ratio, while attenuation can change the order before that regime is reached.

Parametric driving removes this protection. The leading operator $T_i$ is periodic in $\psi=2\nu t$, giving, for a nonzero denominator,
\begin{equation}
 R_{\rm tr}(t)=\mathcal R_{\rm tr}(\psi)+\ord(e^{-\kappa t}),\qquad
 \mathcal R_{\rm tr}(\psi)=\frac{\|T_A(\psi)\|_1}{\|T_B(\psi)\|_1}.\label{eq:ratio}
\end{equation}
Its minimum $\mathcal R_{\rm tr}^-$ and maximum $\mathcal R_{\rm tr}^+$ over $\psi$ fix the eventual order: given $D_A(0)>D_B(0)$, $\mathcal R_{\rm tr}^+<1$ implies a lasting reversal, $\mathcal R_{\rm tr}^->1$ implies eventual nonreversal, and $\mathcal R_{\rm tr}^-<1<\mathcal R_{\rm tr}^+$ implies persistent alternation. The same criterion applies to other diagnostics with a continuous periodic or quasiperiodic leading ratio; Appendix~\ref{app:tail} gives the conditions for this extension. We write $\mathcal R_X^\pm$ for its phase extrema. The isolated real mode of Ref.~\cite{Salazar2026} and the undriven oscillator give constant ratios, for which the extrema coincide. At $n_B=0.33$ the trace interval is $[0.810,0.952]$ [Fig.~\ref{fig:one}(b) and Table~\ref{tab:three}]: the leading ratio keeps oscillating but stays below unity.

The width of this interval quantifies the onset of phase dependence. For $n_A=n_B$, opposite real preparation phases, and $d=\kappa/(2\omega)$, a weak drive gives
\begin{equation}
 \mathcal R_{\rm tr}^+-\mathcal R_{\rm tr}^-=\frac{2(4\sqrt2-2)d}{\sqrt{1+d^2}}\frac{|\xi|}{\omega}
 +\ord(\xi^2/\omega^2).\label{eq:width}
\end{equation}
The full trace-norm width agrees with this result to $0.5\%$ at $\xi=10^{-3}\omega$ and $5\%$ at $\xi=10^{-2}\omega$ (Appendix~\ref{app:tail}). It closes continuously at zero drive: parametric driving exposes the competition hidden by number symmetry. A pump phase $\chi$, introduced through $\xi(e^{i\chi}a^{\dagger2}+e^{-i\chi}a^2)$, makes the results depend only on $\varphi-\chi$. In Fig.~\ref{fig:one}(c), varying $\varphi_A$ redistributes preparation A between the real mode and the complex-conjugate pair while the generator stays fixed. The initial ratio remains above unity, so the crossings of $\mathcal R_{\rm tr}^-$ and $\mathcal R_{\rm tr}^+$ through unity separate eventual nonreversal, persistent alternation, and lasting reversal.

Because Eq.~\eqref{eq:T} depends only on the initial covariance, covariance-matched non-Gaussian states have the same leading tail. An explicit example mixes the vacuum, with weight $1-n_i$, and the superposition $(\ket0+e^{i\varphi_i}\ket2)/\sqrt2$, with weight $n_i$. For the principal pair, the trace-distance ratio of these states differs from the Gaussian one at finite times but converges to the same leading ratio (Appendix~\ref{app:gaussian}).

\section{Opposite lasting orders}\label{sec:orders}
Let $\mu=\sqrt{\det W}>1/2$ be the symplectic eigenvalue of the stationary state, $g_\mu=\log[(\mu+1/2)/(\mu-1/2)]$, and $\Lambda_i=W^{-1/2}\Delta_iW^{-1/2}$. The Gaussian entropy tail is
\begin{align}
 \mathscr S_i(t)&=e^{-2\kappa t}\mathcal Q_W(\Lambda_i(t))+\ord(e^{-3\kappa t}),\label{eq:RE}\\
 \mathcal Q_W(\Lambda)&=\frac{\mu^2(\Tr\Lambda)^2}{8(\mu^2-1/4)}
 +\frac{\mu g_\mu}{8}\big[(\Tr\Lambda)^2-4\det\Lambda\big].\label{eq:Q}
\end{align}
The first (radial) term measures changes of the symplectic eigenvalue; the second (shape) term measures squeezing relative to $W$ at fixed symplectic eigenvalue. The leading entropy ratio is $\mathcal R_{\mathscr S}(\psi)=\mathcal Q_W(\Lambda_A)/\mathcal Q_W(\Lambda_B)$, and its extrema $\mathcal R_{\mathscr S}^\pm$ follow from this closed form. The HS ratio has the analogous closed form given in Appendix~\ref{app:entropy}.

Figure~\ref{fig:one}(d) shows the trace and entropy intervals together with the initial ratios as $n_B$ varies. Opposite trace-distance and relative-entropy orders occupy the range
\begin{equation}
 0.3124<n_B<0.3371,\label{eq:region2}
\end{equation}
where trace distance gives a lasting reversal and relative entropy eventually places A farther. Both diagnostics agree that A is initially farther throughout this range. Table~\ref{tab:three} lists the three orders at $n_B=0.33$. The trace and entropy intervals lie on opposite sides of unity, with margins of $4.8\%$ and $4.2\%$.

\begin{table}[!htbp]
\caption{Initial ratio, leading interval, and eventual order of each diagnostic $X$ for the principal pair, $n_B=0.33$. All three diagnostics initially place A farther.}\label{tab:three}
\begin{ruledtabular}
\begin{tabular}{llll}
$X$ & $R_X(0)$ & $[\mathcal R_X^-,\mathcal R_X^+]$ & Late-time order\\\hline
Trace & 1.0636 & $[0.8100,0.9521]$ & Lasting reversal \\
Entropy & 1.0870 & $[1.0421,1.3847]$ & Eventual nonreversal \\
HS & 1.0728 & $[0.8801,1.0242]$ & Persistent alternation

\end{tabular}
\end{ruledtabular}
\end{table}

Figure~\ref{fig:one}(a) shows how the three orders emerge from the same initial order: the trace-distance ratio stays below unity after $\omega t=1.36$, the entropy ratio returns above unity at $\omega t=2.62$, and the HS ratio keeps crossing unity. The leading ratios in Fig.~\ref{fig:one}(b) explain their persistence. All three oscillate, but only the HS interval straddles unity; a periodic ratio need not imply repeated reversals.

The neighboring ranges show how these orders change with preparation. For $0.2598<n_B<0.3124$, trace distance alternates while relative entropy eventually places A farther. For $0.3371<n_B<0.3536$, trace distance gives a lasting reversal while entropy alternates; the latter endpoint is the equality of the initial entropies. The HS distance alternates for $0.2865<n_B<0.3387$. Each classification is a strict inequality on the phase circle, so continuity extends it to an open region of Gaussian preparations and elliptic quadratic-generator parameters; Appendix~\ref{app:entropy} gives the boundaries, the proof, and a second generator. The eventual order is determined by the phase extrema, but its onset must be found from the full dynamics: the exact entropy ratio approaches its leading expression only after early crossings that the tail does not reproduce (Appendix~\ref{app:entropy}).

The origin of the disagreement becomes clearer when energy and squeezing deviations are separated. The effective energy obeys
\begin{equation}
 \cL_0^\dagger H_\xi=-\kappa H_\xi,\qquad \langle H_\xi\rangle_{\rm ss}=0,\label{eq:energy}
\end{equation}
so its ratio is the constant $|E_A/E_B|$, where $E_i=n_i[\omega+\sqrt2\xi\cos\varphi_i]$. It equals $1.1397$ for the principal pair. The energy deviation measures only the weight of the real mode, whose left eigenoperator is $H_\xi$, whereas the trace distance retains all three modes. Observable-based detection accordingly depends on the measured mode weights~\cite{Bagui2026}.

\begin{figure*}[t]
\includegraphics[width=\textwidth]{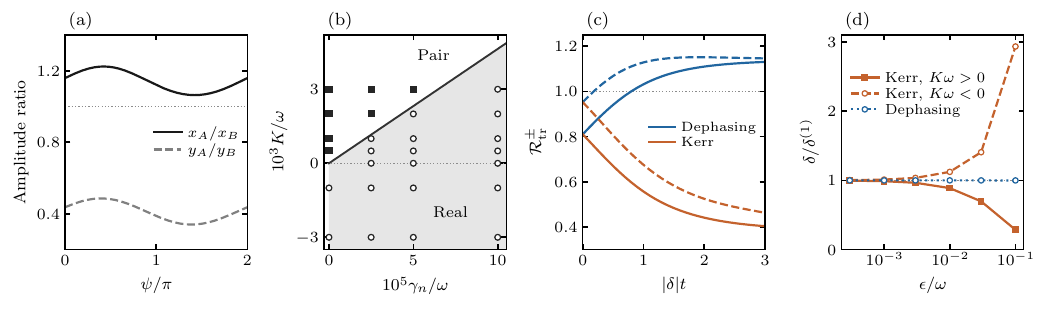}
\caption{Diagnostic weights and selection by weak perturbations for Eq.~\eqref{eq:family}.
(a) Radial-amplitude ratio $x_A/x_B$ (solid) and shape-amplitude ratio $y_A/y_B$ (dashed) for $n_B=0.33$, where $x_i=\Tr\Lambda_i$ and $y_i^2=(\Tr\Lambda_i)^2-4\det\Lambda_i$. A has the larger radial deviation and the smaller shape deviation throughout the cycle.
(b) Simultaneous number dephasing and Kerr interaction. The solid line is the first-order boundary $K/\omega=46.4\gamma_n/\omega$; the shaded side favors the real branch. Full-Liouvillian calculations select the real branch (open circles) or the conjugate pair (filled squares).
(c) Phase extrema $\mathcal R_{\rm tr}^-$ (solid) and $\mathcal R_{\rm tr}^+$ (dashed) at fixed slow time $|\delta|t$ for $n_B=0.33$, with dephasing $\gamma_n=10^{-3}\omega$ (blue) or Kerr interaction $K=10^{-3}\omega$ (orange), each acting alone.
(d) Full splitting divided by its first-order prediction at Fock cutoff $N=34$. Same-sign Kerr selects the pair (filled orange squares), opposite-sign Kerr selects the real branch (open orange circles), and dephasing selects the real branch (open blue circles). Here $\epsilon=|K|$ or $\gamma_n$ as appropriate.}\label{fig:two}
\end{figure*}

In the coordinates set by the stationary covariance, A has the larger radial deviation but the smaller shape deviation throughout the leading cycle [Fig.~\ref{fig:two}(a) and Appendix~\ref{app:entropy}]. The nearly pure stationary state, $\mu=0.5097$, strongly enhances the radial term of Eq.~\eqref{eq:Q}: it supplies more than $91\%$ of $\mathcal Q_W$ for each preparation. Since $W^{-1}\simeq(2/\omega)\operatorname{diag}(\omega+2\xi,\omega-2\xi)$, $\Tr\Lambda_i$ approximately follows the energy deviation, and the entropy keeps A farther. Its ratio lies around the squared energy ratio, $1.2989$, with the residual phase and shape contributions setting the endpoints. The trace distance also responds to the rotating squeezing deviations and places A closer. The squared HS tail gives the radial and shape terms a coefficient ratio of $2$, compared with $11.2$ for relative entropy; for this pair, the resulting balance changes with phase and the HS order alternates.

Classical diagnostics of the same covariance dynamics separate these two mechanisms. The Gaussian Wigner functions are positive and obey an Ornstein--Uhlenbeck equation with the same drift and diffusion (Appendix~\ref{app:classical}). Their total-variation distance from the stationary density shares the trace-distance order: at $n_B=0.33$ its initial ratio is $1.0186$ and its leading interval is $[0.8149,0.9487]$, a lasting reversal, and at $n_B=0.30$ both alternate. The Kullback--Leibler (KL) divergence of the same densities instead weights the radial and shape terms equally; both coefficients in Eq.~\eqref{eq:Q} approach this classical value $1/8$ for $\mu\gg1$. Near purity, $\mu\to1/2$, the radial coefficient $\mu^2/[8(\mu^2-1/4)]$ diverges, whereas the shape coefficient grows only logarithmically; at our parameters their ratio is $11.2$. The KL leading interval is $[0.6032,0.8358]$ and its initial ratio is $0.8617$, so it places A closer from the start. The trace-distance order therefore has a classical counterpart, whereas the entropy order reflects the quantum weighting near a pure stationary state.

A thermal bath provides a direct way to vary these weights. Replacing vacuum damping by $\kappa[(n_{\rm th}+1)\cD[a]+n_{\rm th}\cD[a^\dagger]]$ gives stationary covariance $W_{\rm th}=(2n_{\rm th}+1)W$. With the initial preparations fixed and the reference updated, $n_{\rm th}=0.05$ raises the symplectic eigenvalue to $0.5607$. The entropy interval then becomes $[0.8276,1.1383]$, while its initial ratio is already $0.979<1$ (Table~\ref{tab:SMthermal}); the leading entropy ratio still straddles unity, but the common initial order of the principal example has been lost. The trace interval barely changes to $[0.8042,0.9498]$. Retuning $n_B$ preserves the region of opposite orders only up to $n_{\rm th}\simeq0.0292$ along its continuation from the vacuum bath (Appendix~\ref{app:thermal}).

The common decay rate sustains these distinct orders within quadratic dynamics. A weak perturbation that splits it eventually selects one branch. If the real branch dominates, the diagnostics recover a common order after sufficient mode separation. The quadratic classifications then describe an intermediate window, after the faster modes have decayed and before the splitting changes the strict inequalities. This window closes after a time of order $|\delta|^{-1}$, up to preparation-dependent prefactors, where $\delta$ is the splitting defined in Eq.~\eqref{eq:rule}. We next determine this selection and then compare its time scale with the measurement accuracy.

\section{First-order selection by weak perturbations}\label{sec:selection}

\begin{table}[!htbp]
\caption{Branch selection for the main parameters. Each perturbation acts alone with strength $\epsilon=10^{-3}\omega$, with $|K|=\epsilon$ in the Kerr rows. Here $\delta^{(1)}$ is from Eq.~\eqref{eq:rule}, and $\delta$ is from the full Liouvillian. The thermal result obeys $|\delta/\epsilon|<10^{-10}$.}\label{tab:pert}
\begin{ruledtabular}
\begin{tabular}{lrrl}
Perturbation&$\delta^{(1)}/\epsilon$&$\delta/\epsilon$&Leading\\\hline
$\cD[n]$&2.1250&2.1250&Real\\
$\cD[a^2]$&1.0821&1.0820&Real\\
$\cD[a]+\cD[a^\dagger]$&0&0&Unsplit\\
Kerr, $K\omega>0$&$-0.0458$&$-0.0453$&Pair\\
Kerr, $K\omega<0$&0.0458&0.0464&Real
\end{tabular}
\end{ruledtabular}
\end{table}

We take $\cL=\cL_0+\mathcal V$, where $\mathcal V$ is a weak Lindblad perturbation of strength $\epsilon$ that commutes with the parity superoperator $\rho\mapsto\Pi\rho\Pi$. Parity-invariant preparations isolate the even relaxation sector. The same sector can be accessed through the projected state $\mathcal P_\Pi(\rho)=(\rho+\Pi\rho\Pi)/2$. A displacement introduces an odd component decaying at approximately $\kappa/2$, which can dominate before the even branches separate. The parity projection commutes with the generator and removes this component.

Let $\varrho_j$ and $Y_j$ be right and left eigenoperators of the triple~\eqref{eq:triple}, normalized so that $\Tr(Y_i^\dagger\varrho_j)$ is $1$ for $i=j$ and $0$ otherwise. Define the perturbation matrix
\begin{equation}
 M_{ij}=\Tr[Y_i^\dagger\mathcal V(\varrho_j)],\qquad
 \Sigma_{\mathcal V}=\Tr M.\label{eq:M}
\end{equation}
Let $\lambda_r$ and $\lambda_\pm$ denote the perturbed eigenvalues continued from $-\kappa$ and $-\kappa\pm2i\nu$. Their first-order shifts $\lambda_j^{(1)}=M_{jj}$ sum to $\Sigma_{\mathcal V}$, and Hermiticity preservation makes $\lambda_\pm^{(1)}$ conjugate and $\lambda_r^{(1)}$ real. Consequently,
\begin{equation}
 \delta\equiv\lambda_r-\operatorname{Re}\lambda_+
 =\underbrace{\frac{3\lambda_r^{(1)}-\Sigma_{\mathcal V}}2}_{\delta^{(1)}}+\ord(\epsilon^2/\omega),\label{eq:rule}
\end{equation}
for regular branches at fixed $\nu>0$. When this continued triple remains the leading even spectrum, $\delta>0$ selects the real branch and $\delta<0$ selects the pair. The biorthogonal projection also applies when the perturbed dynamics is non-Gaussian.

For number dephasing $\gamma_n\cD[n]$, the closed second-moment equations give $\lambda_r^{(1)}=8\xi^2\gamma_n/\nu^2$ and $\Sigma_{\mathcal V}=-4\gamma_n$. Equation~\eqref{eq:rule} then gives $\delta^{(1)}=2(1+6\xi^2/\nu^2)\gamma_n>0$, which selects the real branch. For the Kerr interaction $H_K=(K/2)n(n-1)$, differentiated Gaussian moments give
\begin{equation}
 \lambda_r^{(1)}=-\alpha K,\quad \Sigma_{\mathcal V}=0,\quad
 \alpha=\frac{24\kappa\omega\xi^2}{\nu^2(\kappa^2+4\nu^2)}.\label{eq:alpha}
\end{equation}
Thus a weak Kerr shift with $K\omega>0$ favors the pair, whereas $K\omega<0$ favors the real branch. The symmetry of Sec.~\ref{sec:model} gives $\delta(-\omega,-K)=\delta(\omega,K)$, relating the two detuning signs exactly.

Two-photon loss $\gamma_2\cD[a^2]$ gives
\begin{equation}
 \delta=\gamma_2\left(1+\frac{6\xi^2}{\nu^2}+\frac{8\xi^2}{\kappa^2+4\nu^2}\right)
 +\ord(\gamma_2^2/\omega)>0.\label{eq:loss2}
\end{equation}
Linear jumps retain equal covariance decay rates. In particular, balanced thermal noise $\gamma_T(\cD[a]+\cD[a^\dagger])$, with $\gamma_T=\kappa n_{\rm th}$, changes only diffusion: it leaves the splitting unchanged but alters the entropy weights through $W_{\rm th}$ (Sec.~\ref{sec:orders}). Table~\ref{tab:pert} compares the first-order rule with the full Liouvillian.

For simultaneous dephasing and Kerr interaction, the first-order boundary in Fig.~\ref{fig:two}(b) is $K/\omega=46.4\gamma_n/\omega$. Thus dephasing dominates at these parameters unless the same-sign Kerr strength exceeds about $46\gamma_n$; opposite-sign Kerr reinforces real-branch selection. Higher-order terms matter on the boundary (Appendix~\ref{app:selection}).

The selected branch fixes the remaining preparation dependence. If the real branch dominates, $\rho_i-\rho_{\rm ss}=c_i e^{\lambda_rt}\varrho_r+o(e^{\lambda_rt})$. Every norm then gives the ratio $|c_A/c_B|$, and a smooth divergence with a positive quadratic response along $\varrho_r$ gives its square, as follows from Ref.~\cite{Salazar2026} for normalized operator-convex Petz divergences. Since $Y_r\propto H_\xi$ as the perturbation vanishes, $c_i\propto E_i$: every weak perturbation that selects the real branch restores the energy ratio $|E_A/E_B|$ up to $\ord(\epsilon)$ corrections. At strength $10^{-3}\omega$, number dephasing gives the trace ratio $1.1397$, two-photon loss $1.1408$, and opposite-sign Kerr $1.1391$, all above unity. Same-sign Kerr instead selects the pair. Interference between its conjugate eigenoperators keeps the ratio periodic, in the narrow interval $[0.4090,0.4129]$, entirely below unity.

Figure~\ref{fig:two}(c) shows how these orders are selected. As the competing contribution decays, the dephasing interval approaches a fixed ratio above unity, whereas same-sign Kerr retains a narrow interval below it. Plotting against the slow time $|\delta|t$ separates this gradual selection from the rapid phase oscillations. Selection takes a time of order $|\delta|^{-1}$, during which the distances decay roughly as $e^{-\kappa t}$. The remaining absolute distance decides whether the selected order can be measured.

Figure~\ref{fig:two}(d) tests the first-order rule: splitting errors stay below $4\%$ for both Kerr signs at $|K|\leq0.003\omega$, while the dephasing error remains below $0.07\%$ at $\gamma_n=0.1\omega$. Appendix~\ref{app:selection} gives the larger-cutoff spectral checks and the additional slow branch found at stronger drive.

\section{Experimental considerations}\label{sec:experiment}
The detuned parametric amplifier arises in degenerate optical parametric amplification~\cite{Collett1984} and in parametrically driven superconducting resonators, either flux-pumped at twice the resonance frequency~\cite{Beaulieu2025} or pumped near resonance through the Kerr nonlinearity~\cite{Castellanos2008}; in the latter case the rotating frame is at the pump frequency. Because the detuning is set by the pump frequency, $\kappa$ comparable to $\omega$ is readily reached. Trapped-ion squeezing, parametric control, and motional tomography provide another route to the required oscillator preparations and readout~\cite{Kienzler2015,Burd2019Parametric,Leibfried1996Tomography}. For Gaussian evolution, measured first and second moments determine the density operators and hence all three diagnostics. Residual nonlinear dynamics requires additional state information or a validated Gaussian approximation.

The experimental signals are the A--B differences of each diagnostic. At zero bath occupation, the unperturbed trace and entropy trajectories display their eventual orders after the entropy return at $\omega t=2.62$, when $(D_A,D_B)\simeq(0.106,0.117)$. The margins are not uniform in time: the trace ratio reaches $0.9839$ near $\omega t=3.82$, and the entropy ratio falls to $1.0046$ near $5.46$. A measurement at $\omega t=3$ avoids these small margins and gives
\begin{equation}
 D_A-D_B=-7.26\times10^{-3},\qquad
 \mathscr S_A-\mathscr S_B=1.06\times10^{-2}.\label{eq:signal}
\end{equation}
Total absolute errors $e_D\leq2\times10^{-3}$ per trace distance and $e_{\mathscr S}\leq3\times10^{-3}$ per entropy suffice to resolve these opposite signs, including reference-state uncertainty. Linear propagation of covariance errors gives characteristic precision scales of $1.69\times10^{-3}$ for the entropy difference and $2.80\times10^{-3}$ for the trace-distance difference. These local estimates put the required covariance accuracy at a few tenths of a percent of the vacuum variance. Using the same measured reference for A and B substantially reduces its contribution to the entropy-difference error (Appendix~\ref{app:finite}). At $\omega t=3$ the HS distance places A closer ($-2.36\times10^{-3}$) and at $\omega t=4$ farther ($+3.12\times10^{-3}$), so its alternation is visible in the same window. For $\kappa/(2\pi)=1$~MHz, the main parameters give $\omega/(2\pi)=2$~MHz, and $\omega t=3$ corresponds to $0.24~\mu$s.

\begin{figure*}[t]
\includegraphics[width=\textwidth]{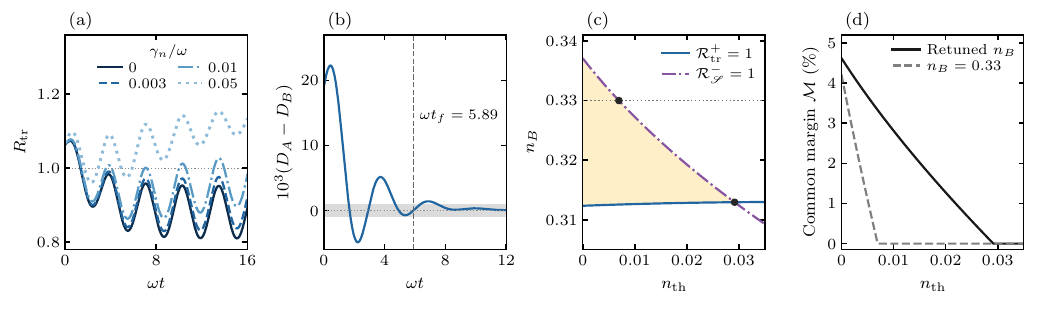}
\caption{Observation window for the principal pair, $n_B=0.33$, with the other parameters of Eq.~\eqref{eq:family}.
(a) Full-Liouvillian trace-distance ratios with a vacuum bath at $\gamma_n/\omega=0$ (solid), $0.003$ (dashed), $0.01$ (dash-dotted), and $0.05$ (dotted).
(b) Scaled distance difference $10^3(D_A-D_B)$ at $\gamma_n=0.05\omega$. The gray band denotes $|D_A-D_B|\leq2e_D$ for $e_D=5\times10^{-4}$ per distance. The vertical dashed line marks the last crossing, $\omega t_f=5.89$, after which A remains farther while both distances decay.
(c) Thermal bath with $\gamma_n=K=\gamma_2=0$ and the stationary reference recomputed at each $n_{\rm th}$: boundaries $\mathcal R_{\rm tr}^+=1$ (blue) and $\mathcal R_{\mathscr S}^-=1$ (purple). The shaded region has opposite lasting orders and both initial ratios above unity; the horizontal dotted line marks $n_B=0.33$. Dots mark the exit of the fixed pair at $n_{\rm th}=0.00695$ and the closure at $(n_{\rm th},n_B)=(0.0292,0.3130)$.
(d) Positive common margin $\mathcal M$, Eq.~\eqref{eq:margin}, with $n_B$ optimized at each $n_{\rm th}$ (solid) or fixed at $0.33$ (dashed).}\label{fig:three}
\end{figure*}

Dephasing limits this observation window. For the principal pair with a vacuum bath, the trajectories computed at $\gamma_n/\omega=0$, $0.001$, $0.003$, and $0.005$ all retain a continuous trace-reversal window covering $2\leq\omega t\leq10$. At $0.005\omega$ (equivalently $\gamma_n/\kappa=0.01$), the first return occurs at $\omega t=16.68$, whereas at $0.01\omega$ it occurs at $3.58$. Figure~\ref{fig:three}(a) shows that a first return above unity can be followed by further crossings before the order settles.

Figure~\ref{fig:three}(b) tests whether the last crossing is visible in an absolute signal. At $\gamma_n=0.05\omega$, the last crossing is at $\omega t=5.89$, with $D_B\simeq0.0224$. At $\omega t=7$, the returned difference is $1.40\times10^{-3}$, resolved by errors below $5\times10^{-4}$ per distance. For $\gamma_n/\omega=0.005$ and $0.01$, the last crossing occurs only after $D_B$ falls below $10^{-8}$; at $0.001$ and $0.003$ it does not occur before $\omega t=80$. The same selected real branch can therefore become visible at accessible times or only after the signal is lost, depending on the dephasing rate. Appendix~\ref{app:finite} gives the full scan. Preparation errors that break parity must be measured or removed by parity projection.

Thermal occupation imposes a separate constraint. With thermal damping alone and fixed $n_B=0.33$, heating weakens the radial enhancement and the entropy minimum reaches unity at $n_{\rm th}=0.00695$, while the trace interval changes little (Appendix~\ref{app:thermal}). Retuning $n_B$ extends the region of opposite orders [Fig.~\ref{fig:three}(c)] with a progressively smaller common margin,
\begin{equation}
 \mathcal M=\min\{1-\mathcal R_{\rm tr}^+,\ \mathcal R_{\mathscr S}^--1\},\label{eq:margin}
\end{equation}
evaluated subject to both initial ratios exceeding unity. Its optimum decreases from $4.6\%$ in a vacuum bath to $2.8\%$ at $n_{\rm th}=0.01$ [Fig.~\ref{fig:three}(d)]. Retuning thus increases thermal tolerance at the cost of a narrower preparation window. The relevant occupation is set by the physical resonator frequency $\omega_c$, rather than the rotating-frame detuning, through $n_{\rm th}=[\exp(\hbar\omega_c/k_BT)-1]^{-1}$. For $\omega_c/2\pi=5$~GHz, the fixed-pair limit $n_{\rm th}=0.00695$ corresponds to $48$~mK and the retuned limit $0.0292$ to $67$~mK; at $45$~mK ($n_{\rm th}=0.005$) the fixed pair still keeps a $1.1\%$ entropy margin. A microwave implementation therefore requires an effective mode temperature below about $48$~mK for the fixed pair, or $67$~mK with retuning, which calls for calibrated thermalization and filtering~\cite{Wang2019Attenuators}. Equilibrium occupation is negligible at optical frequencies.

\section{Conclusions}\label{sec:conclusion}
Observing longer does not by itself settle which preparation is closer: the eventual order depends on the diagnostic. Open sets of Gaussian preparations give a lasting trace-distance Mpemba reversal and the opposite relative-entropy order, with HS distance still alternating. The underlying coincidence of decay rates is structural for single-mode quadratic oscillators with linear damping in the elliptic regime. Number symmetry fixes the undriven trace tail; parametric driving opens phase-dependent intervals. Near a pure stationary state, the relative entropy strongly weights changes of the symplectic eigenvalue, which explains its different order. Weak perturbations split the common decay rate; their strength and the measurement precision determine whether branch selection is visible. Diagnostic choice and observation time are therefore part of the physical characterization of a lasting Mpemba reversal.

\begin{acknowledgments}
The author used OpenAI's ChatGPT to assist with code optimization, preparation of the submission files, and checks of the analytical derivations.
\end{acknowledgments}

\section*{Data availability}
The data and code that support the findings of this article are openly available in Zenodo~\cite{He2026Data}.

\appendix
\onecolumngrid
\section{Conventions and Gaussian preparations}\label{app:gaussian}
We use $\hbar=1$, $a=(Q+iP)/\sqrt2$, $[Q,P]=i$, and
$\cD[L]\rho=L\rho L^\dagger-\{L^\dagger L,\rho\}/2$. The Kerr Hamiltonian is
$H_K=K a^{\dagger2}a^2/2=K n(n-1)/2$. A Fock cutoff $N$ retains $\ket0,\ldots,\ket{N-1}$; analytical formulas refer to the infinite oscillator. $G[V]$ denotes the centered Gaussian state with covariance $V$, a squeezed thermal state with thermal occupation $\sqrt{\det V}-1/2$~\cite{Weedbrook2012}. At $\omega=1$ and the main parameters with $n_B=0.33$,
\begin{equation}
\begin{gathered}
 V_A=\begin{pmatrix}0.646447&0\\0&1.353553\end{pmatrix},\qquad
 V_B=\begin{pmatrix}1.063345&0\\0&0.596655\end{pmatrix},\\[3pt]
 W=\begin{pmatrix}0.421760&-0.024450\\-0.024450&0.617359\end{pmatrix}.
\end{gathered}\label{eq:SMcov}
\end{equation}
The initial covariances are held fixed when a perturbation is added; the reference density operator is then the stationary state of that perturbed generator.

\begin{figure}[ht]
\centering
\includegraphics[width=\textwidth]{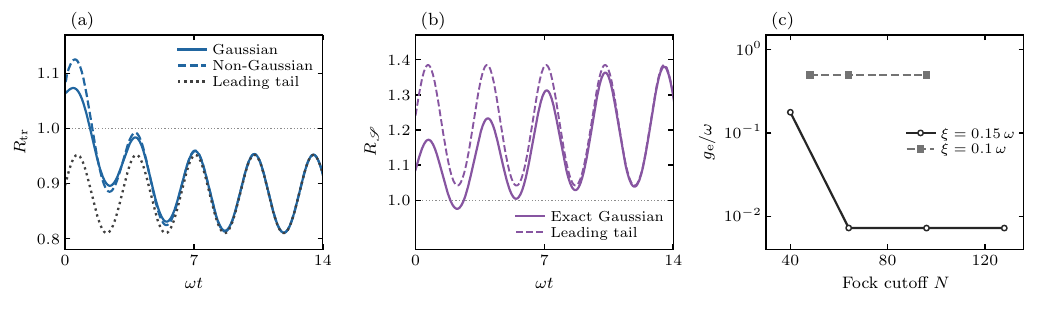}
\caption{Tail convergence and spectral checks. Panels (a,b) use the principal pair, $n_B=0.33$, and the other parameters of Eq.~\eqref{eq:family}.
(a) Finite-time trace ratios for Gaussian preparations (solid) and non-Gaussian preparations with matching initial covariances (dashed), together with their common leading expression (dotted).
(b) Exact Gaussian relative-entropy ratio (solid) and its leading periodic tail (dashed). The curves merge for $\omega t\gtrsim10$; the transient entropy reversal near $\omega t\simeq2$ is absent from the tail. Horizontal dotted lines in (a,b) mark unity. (c) Leading even decay rate $g_{\rm e}=-\max_{\lambda\ne0}\operatorname{Re}\lambda$ versus Fock cutoff $N$ at $K=-0.03\omega$, for $\xi=0.15\omega$ (circles) and $0.1\omega$ (squares). The stronger drive produces an additional slow mode, as discussed in Appendix~\ref{app:selection}.}\label{fig:SMchecks}
\end{figure}

The non-Gaussian mixture of the vacuum (weight $1-n_i$) and the pure state $(\ket0+e^{i\varphi_i}\ket2)/\sqrt2$ (weight $n_i$) has the same covariance. For the principal pair, the initial trace distances are $(0.3011,0.2831)$ for the Gaussian states and $(0.3079,0.2839)$ for the non-Gaussian states. Figure~\ref{fig:SMchecks}(a) shows that both trace ratios approach the common leading expression as the additional higher-moment contributions decay.

\section{Undriven symmetry protection and the driven tail}\label{app:tail}
\subsection{Vacuum attenuation}
At $\xi=0$, number rotation commutes with vacuum attenuation. For a parity-invariant finite-Fock input,
\begin{equation}
\begin{aligned}
 \rho_i(t)-\ket0\bra0
 &=e^{-\kappa t}\Big[-n_i\ket0\bra0+n_i\ket1\bra1\\
 &\qquad+\frac{m_ie^{-2i\omega t}}{\sqrt2}\ket2\bra0
 +\frac{m_i^*e^{2i\omega t}}{\sqrt2}\ket0\bra2\Big]
 +\ord(e^{-2\kappa t}).
\end{aligned}
\label{eq:SMvacuum}
\end{equation}
Differentiating the Gaussian characteristic function gives the same expansion for a centered Gaussian input. The nonzero eigenvalues of the coefficient are $n_i$ and $(-n_i\pm\sqrt{n_i^2+2|m_i|^2})/2$, so the leading distance is $e^{-\kappa t}C_i$ with $C_i$ given by Eq.~\eqref{eq:protected}, independent of rotation phase. Attenuation can produce finite-time crossings and a lasting reversal even though the asymptotic ratio $C_A/C_B$ is constant. Exact time-independent proportionality holds only when $\rho_A-\rho_{\rm ss}=c_0U_\theta(\rho_B-\rho_{\rm ss})U_\theta^\dagger$, with real $c_0$ and $U_\theta=e^{-i\theta n}$.

\subsection{Gaussian channel and uniform asymptotics}
For $\Phi(\bm k)=\Tr[\rho e^{i\bm k^T\bm r}]$, the quadratic channel gives
\begin{equation}
 \Phi_i(t,\bm k)=\Phi_i(0,e^{\mathsf F^Tt}\bm k)
 \exp[-\tfrac12\bm k^T(W-e^{\mathsf F t}We^{\mathsf F^Tt})\bm k].
\end{equation}
Parity removes the odd powers in the expansion at $\bm k=0$, and the leading terms are
\begin{equation}
 \Phi_i(t,\bm k)=e^{-\bm k^TW\bm k/2}
 [1-\tfrac12 e^{-\kappa t}\bm k^T\Delta_i(t)\bm k+\ord(e^{-2\kappa t})].
\end{equation}
The quadratic term is represented by $T_i(t)$ of Eq.~\eqref{eq:T}. For nonzero drive the stationary covariance lies in the interior of the physical domain and derivatives of $G[W]$ are trace class. Gaussian inputs and finite-Fock inputs (whose characteristic functions are polynomials times a Gaussian) therefore give a trace-norm remainder uniform over the compact phase circle. At zero drive Eq.~\eqref{eq:SMvacuum} handles the vacuum boundary directly.

Since $S(t+\pi/\nu)=-S(t)$, $T_i$ has period $\pi/\nu$. If $\min_\psi\|T_B(\psi)\|_1>0$, uniform convergence of the ratio follows from the triangle inequality. A strict interval crossing of unity then gives crossings at arbitrarily late times; a strict one-sided interval fixes the eventual ordering. Equality at an endpoint is a boundary requiring subleading terms.

The argument extends to a diagnostic with $X_i(t)=e^{-\kappa_X t}[f_i(\bm\theta(t))+o(1)]$, where the leading decay rate $\kappa_X>0$ is common to both preparations, the remainder is uniform in phase, and $f_i$ is continuous. For the quadratic oscillator considered here, $\kappa_X=\kappa$ for the trace and HS norms and $\kappa_X=2\kappa$ for relative entropy and squared norms. If $f_B$ has a positive lower bound, the leading ratio is $f_A/f_B$. Its extrema are taken over the closure of the periodic or quasiperiodic phase orbit. Recurrence then gives persistent alternation when the extrema straddle unity, while a strict one-sided interval fixes the eventual order. A finite diagonalizable leading spectral sector, separated from faster modes, supplies this form for norms and for smooth divergences with a nonzero leading homogeneous response.

The stationary moments are
\begin{equation}
 n_{\rm ss}=\frac{8\xi^2}{\Gamma},\qquad
 m_{\rm ss}=-\frac{2\xi(2\omega+i\kappa)}{\Gamma},\qquad
 \Gamma=\kappa^2+4\nu^2.
 \label{eq:SMmoments}
\end{equation}
For linear jumps $L_j=u_j a+v_j a^\dagger$, the jump contributions to the first-moment equation are $-\frac12\sum_j(|u_j|^2-|v_j|^2)\langle a\rangle$, with no term in $\langle a^\dagger\rangle$. The drift is therefore $\mathsf F_H-\kappa_{\rm eff}I/2$, and for $\kappa_{\rm eff}>0$ the covariance eigenvalues are $-\kappa_{\rm eff}$ and $-\kappa_{\rm eff}\pm2i\nu$ in the elliptic regime. Diffusion determines $W$ and the norm weights, while these three real parts remain equal.

\subsection{Weak-drive interval width}
Define $\bar\xi=\xi/\omega$, $q=n_A/n_B$, and $d=\kappa/(2\omega)$. Opposite real phases $(\varphi_A,\varphi_B)=(\pi,0)$ give
\begin{align}
 \mathcal R_{\rm tr}(\psi)&=q-\bar\xi(b_1+b_2)
 +\bar\xi b_3\cos(\psi-\psi_1)+\ord(\bar\xi^2),\\
 b_1&=q(4\sqrt2-2),\qquad
 b_2=\frac{q(\sqrt2-1)(n_A^{-1}+n_B^{-1})}{1+d^2},\nonumber\\
 b_3&=\frac{b_1d}{\sqrt{1+d^2}},\qquad \psi_1=\arctan(1/d).\nonumber
\end{align}
These coefficients follow by differentiating $W$, $S$, and the Hermitian trace norm on the nonzero support of Eq.~\eqref{eq:SMvacuum}; the stationary-covariance change contributes $b_2$. At $q=1$ the width $2|\bar\xi|b_3$ is Eq.~\eqref{eq:width}. For $n_A=n_B=0.5$ and $\kappa=0.5\omega$, the full trace-norm width deviates from it by $-0.52\%$, $-5.2\%$, and $-20\%$ at $\bar\xi=10^{-3}$, $10^{-2}$, and $0.04$, a relative correction $\simeq-5.2\,\bar\xi$ at weak drive. A pump phase rotates the covariance axes, so the results depend only on the preparation phases measured from the pump phase.

\section{Relative entropy, classification boundaries, and openness}\label{app:entropy}
\subsection{Exact entropy and its quadratic coefficient}
Define $s(\mu)=(\mu+1/2)\log(\mu+1/2)-(\mu-1/2)\log(\mu-1/2)$. For a full-rank $G[W]$, the Gibbs matrix is $\mu g_\mu W^{-1}$ with natural logarithms, and the Gaussian relative entropy is~\cite{Pirandola2017}
\begin{equation}
 \mathscr S(G[V]\Vert G[W])=s(\mu)-s(\sqrt{\det V})
 +\frac{\mu g_\mu}{2}\Tr[W^{-1}(V-W)].\label{eq:SMREexact}
\end{equation}
For $V=W+\Delta$ and $\Lambda=W^{-1/2}\Delta W^{-1/2}$,
\begin{equation}
 \sqrt{\det V}=\mu\left[1+\tfrac12\Tr\Lambda
 +\tfrac12\det\Lambda-\tfrac18(\Tr\Lambda)^2\right]+\ord(\|\Delta\|^3).
\end{equation}
Using $s'(\mu)=g_\mu$ and $s''(\mu)=-(\mu^2-1/4)^{-1}$ yields $\mathscr S(G[W+\Delta]\Vert G[W])=\mathcal Q_W(\Lambda)+\ord(\|\Delta\|^3)$, with $\mathcal Q_W$ given by Eq.~\eqref{eq:Q}. Both terms are nonnegative because $\Lambda$ is symmetric. Equivalently, $\mathcal Q_W$ is one half of the Bogoliubov--Kubo--Mori metric~\cite{Petz1996} evaluated on the linearized density operator $T$ of Eq.~\eqref{eq:T}. Finite-time entropies are computed from Eq.~\eqref{eq:SMREexact} with 65-digit arithmetic to control cancellation near stationarity.

For the HS norm, independent Gaussian formulas give
\begin{align}
 \|G[V]-G[W]\|_2^2&=\frac1{2\sqrt{\det V}}+\frac1{2\sqrt{\det W}}
 -\frac2{\sqrt{\det(V+W)}},\\
 \|T\|_2^2&=\frac{3(\Tr\Lambda)^2-4\det\Lambda}{32\mu}.
\end{align}

\subsection{Entropy weights near the purity boundary}
Put $x=\Tr\Lambda$ and $y^2=(\Tr\Lambda)^2-4\det\Lambda$. The first-order change of the symplectic eigenvalue in direction $\Delta$ is $\mu x/2$. Thus
\begin{equation}
 \mathcal Q_W=a_{\rm rad}x^2+a_{\rm sh}y^2,\qquad
 a_{\rm rad}=\frac{\mu^2}{8(\mu^2-1/4)},\qquad
 a_{\rm sh}=\frac{\mu g_\mu}{8}.
\end{equation}
The radial term measures the change of symplectic eigenvalue; the shape term measures the eigenvalue difference of $\Lambda$. As $\mu\downarrow1/2$, the coefficients diverge algebraically and logarithmically, respectively. At the main parameters, $\mu=0.50969$, $a_{\rm rad}=3.3203$, $a_{\rm sh}=0.29604$, and $a_{\rm rad}/a_{\rm sh}=11.216$. Over the phase circle, the radial fractions $h_i=a_{\rm rad}x_i^2/\mathcal Q_{W,i}$ satisfy $h_A\in[0.9847,0.9902]$ and $h_B\in[0.9104,0.9118]$, while $x_A/x_B\in[1.0638,1.2238]$. Since $\mathcal Q_{W,A}/\mathcal Q_{W,B}=(x_A/x_B)^2h_B/h_A$,
\begin{equation}
 \frac{\mathcal Q_{W,A}}{\mathcal Q_{W,B}}\geq
 \left[\min_\psi\frac{x_A}{x_B}\right]^2
 \frac{\min_\psi h_B}{\max_\psi h_A}\simeq1.0405>1.
\end{equation}
The coefficient ratio alone does not give the ratio of contributions: the radial-to-shape ratio ranges from $64$ to $101$ for A and from $10.2$ to $10.3$ for B, and keeping only the radial terms overestimates the entropy ratio by $8.2$--$8.6\%$.

Removing the entropy coefficients gives $x_A^2/y_A^2\in[5.730,9.027]$ and $x_B^2/y_B^2\in[0.906,0.922]$. With $y_i\geq0$, the shape-amplitude ratio is $y_A/y_B\in[0.34014,0.48635]$: A has the larger radial deviation but the smaller shape deviation throughout the cycle. In these variables the HS coefficient is $\|T\|_2^2=(2x^2+y^2)/(32\mu)$, while the classical KL coefficient is $(x^2+y^2)/8$. Their radial-to-shape coefficient ratios are $2$ and $1$, respectively, compared with $11.216$ for relative entropy at the main parameters.

The energy connection follows from the exact identity
\begin{equation}
 W^{-1}=2I+\frac{8\xi}{\kappa^2+4\omega^2}
 \begin{pmatrix}2\omega&\kappa\\\kappa&-2\omega\end{pmatrix}
 \simeq\frac2\omega\operatorname{diag}(\omega+2\xi,\omega-2\xi),
\end{equation}
where the approximation is for small $\kappa/\omega$ at fixed $\xi/\omega$. It gives $x_i\simeq4E_i/\omega$, with $E_i$ defined below Eq.~\eqref{eq:energy}. Energy equality occurs at $n_B=0.3761$, inside the entropy-alternating range.

\subsection{Preparation intervals and initial orders}
For the main parameters, periodic extrema are refined over all detected local extrema, and a scalar root solver locates each boundary in the scan $0.16\leq n_B\leq0.42$ (spacing $0.0005$). Table~\ref{tab:SMboundaries} lists the roots. Table~\ref{tab:SMexamples} gives interior points on either side of the principal pair and a representative point of a second generator.
\begin{table}[ht]
\caption{Preparation boundaries for the main parameters, including each diagnostic's initial-order equality. ``Farther'' and ``closer'' refer to the late-time position of A.}\label{tab:SMboundaries}
\begin{ruledtabular}\begin{tabular}{lll}
Condition & $n_B$ & Change as $n_B$ increases\\\hline
$\mathcal R_{\rm tr}^-=1$ & 0.259753 & Trace: farther to alternating \\
$\mathcal R_{\rm HS}^-=1$ & 0.286532 & HS: farther to alternating \\
$\mathcal R_{\rm tr}^+=1$ & 0.312353 & Trace: alternating to closer \\
$\mathcal R_{\mathscr S}^-=1$ & 0.337102 & Entropy: farther to alternating \\
$\mathcal R_{\rm HS}^+=1$ & 0.338709 & HS: alternating to closer \\
$\mathscr S_A(0)=\mathscr S_B(0)$ & 0.353614 & Initial entropy order changes \\
$D_A(0)=D_B(0)$ & 0.370049 & Initial trace order changes \\
$D_{{\rm HS},A}(0)=D_{{\rm HS},B}(0)$ & 0.378071 & Initial HS order changes \\
$\mathcal R_{\mathscr S}^+=1$ & 0.388939 & Entropy: alternating to closer

\end{tabular}\end{ruledtabular}
\end{table}

\begin{table}[ht]
\caption{Interior Gaussian preparations for two quadratic generators, both with $(n_A,\varphi_A)=(0.5,\pi)$ and $\varphi_B=0$; Table~\ref{tab:three} gives $n_B=0.33$. Each initial and late-time comparison uses the same diagnostic.}\label{tab:SMexamples}
\begin{ruledtabular}\begin{tabular}{ccll}
$n_B$ & Diagnostic & Initial ratio & Late-time interval\\\hline
\multicolumn{4}{l}{$\xi=0.1\omega,\ \kappa=0.5\omega$} \\
0.30 & Trace & 1.1202 & $[0.8816,1.0365]$ \\
 & Relative entropy & 1.2200 & $[1.2532,1.6718]$ \\
 & Hilbert--Schmidt & 1.1284 & $[0.9595,1.1174]$ \\
0.35 & Trace & 1.0303 & $[0.7684,0.9030]$ \\
 & Relative entropy & 1.0125 & $[0.9297,1.2325]$ \\
 & Hilbert--Schmidt & 1.0404 & $[0.8340,0.9702]$ \\
\hline\multicolumn{4}{l}{$\xi=0.15\omega,\ \kappa=0.5\omega$} \\
0.255 & Trace & 1.0872 & $[0.7779,0.9512]$ \\
 & Relative entropy & 1.1269 & $[1.0456,1.5821]$ \\
 & Hilbert--Schmidt & 1.0976 & $[0.8644,1.0518]$

\end{tabular}\end{ruledtabular}
\end{table}

Parameterize an elliptic quadratic generator and two physical centered Gaussian covariances by $\bm p$. Where the stationary covariance is full rank and the denominator stays positive, the leading ratio $\mathcal R(\bm p,\psi)$ is jointly continuous. Compactness of the phase circle makes its extrema continuous in $\bm p$. Strict one-sided order, strict alternation, and the initial inequality therefore survive sufficiently small changes of $\bm p$, which proves the open-region statement of Sec.~\ref{sec:orders}. As an independent example, the generator $\xi=0.15\omega$, $\kappa=0.5\omega$ has the opposite-order range $0.2398<n_B<0.2614$; its point $n_B=0.255$ is in Table~\ref{tab:SMexamples}. 

The entropy sign at $n_B=0.33$ is also certified uniformly in phase. Since $\Delta(\psi)$ has harmonics $0,\pm1$, the difference $\Delta\mathcal Q(\psi)=\mathcal Q_W(\Lambda_A)-\mathcal Q_W(\Lambda_B)$ is a trigonometric polynomial of degree two. Bounding its derivative by the sum of the harmonic amplitudes, $4096$ equally spaced samples give $\min_\psi\Delta\mathcal Q(\psi)\geq0.3862$. Figure~\ref{fig:SMchecks}(b) links this asymptotic bound to the full evolution: the exact and leading entropy ratios agree at late times, whereas their early-time difference explains why a transient entropy reversal can precede eventual nonreversal.

\section{Classical total variation and KL divergence}\label{app:classical}
For normalized Wigner densities on phase space $\bm q$, let $D_{{\rm TV},i}=\frac12\int|f_i-f_W|\,d^2q$. The positive Wigner densities of the Gaussian preparations obey the Ornstein--Uhlenbeck equation
\begin{equation}
 \partial_tf=-\nabla\cdot(\mathsf F\bm q f)+\frac\kappa4\nabla^2 f,
\end{equation}
with the same covariance dynamics. A small covariance displacement $\Delta$ gives
\begin{equation}
 f_{W+\Delta}-f_W=\frac12 f_W
 [\bm q^TW^{-1}\Delta W^{-1}\bm q-\Tr(W^{-1}\Delta)]+\ord(\|\Delta\|^2).
\end{equation}
Whitening $\bm q=W^{1/2}\bm u$ yields the total-variation coefficient
\begin{equation}
 C_{\rm TV}(\Lambda)=\frac14\mathbb E\left|\bm u^T\Lambda\bm u-\Tr\Lambda\right|,
 \qquad \bm u\sim\mathcal N(0,I),
\end{equation}
which we evaluate by an exact radial integral and angular quadrature. The principal leading interval is $[0.8149,0.9487]$. Direct integration of the initial Wigner densities of the Gaussian preparations gives $D_{{\rm TV},A}(0)=0.22497$ and $D_{{\rm TV},B}(0)=0.22087$, a ratio of $1.0186>1$. Combined with the strict leading interval below unity, this establishes a lasting reversal in classical total variation as well.

The KL divergence of the same centered normal densities is
\begin{align}
 D_{\rm KL}(f_V\Vert f_W)&=\tfrac12\left[\Tr(W^{-1}V)-2-\log\frac{\det V}{\det W}\right]\nonumber\\
 &=\tfrac14\Tr(\Lambda^2)+\ord(\|\Delta\|^3).
\end{align}
In the variables of Appendix~\ref{app:entropy}, the coefficient is $(x^2+y^2)/8$: radial and shape contributions have equal weights. At $n_B=0.33$, the initial KL values are $(0.2597,0.3014)$, with ratio $0.8617$, and the leading interval is $[0.6032,0.8358]$. Both comparisons place A closer. Thus the initial KL order differs from that of the three quantum diagnostics.

\section{Thermal occupation of the reservoir}\label{app:thermal}
We replace the vacuum dissipator by a thermal reservoir with occupation $n_{\rm th}$,
\begin{equation}
 \cL_{\rm th}\rho=-i[H_\xi,\rho]+\kappa(n_{\rm th}+1)\cD[a]\rho
 +\kappa n_{\rm th}\cD[a^\dagger]\rho.
\end{equation}
The drift $\mathsf F$ is unchanged, while the diffusion is $\kappa(n_{\rm th}+1/2)I$. Hence
\begin{equation}
 W_{\rm th}=(2n_{\rm th}+1)W,\qquad \mu_{\rm th}=(2n_{\rm th}+1)\mu,
\end{equation}
where $W$ and $\mu$ refer to the vacuum bath. Initial covariances remain fixed, and every diagnostic uses $G[W_{\rm th}]$ as its reference. The formulas of Appendix~\ref{app:entropy} apply with $W\to W_{\rm th}$ and $\Delta_i=S(V_i-W_{\rm th})S^T$. The energy deviation obeys $\langle H_\xi\rangle-\omega n_{\rm th}\propto e^{-\kappa t}$, and the covariance eigenvalues remain $-\kappa,-\kappa\pm2i\nu$. Thermal occupation changes the weights and preparation overlaps while retaining equal decay rates.

\begin{table}[ht]
\caption{Thermal reservoir at fixed $n_B=0.33$ and otherwise the main parameters. Initial states are held fixed and the stationary reference is recomputed.}\label{tab:SMthermal}
\begin{ruledtabular}\begin{tabular}{rrrrrll}
$n_{\rm th}$ & $\mu_{\rm th}$ & $a_{\rm rad}/a_{\rm sh}$ & $R_{\rm tr}(0)$ & $R_{\mathscr S}(0)$ & Trace interval & Entropy interval\\\hline
0 & 0.5097 & 11.22 & 1.0636 & 1.0870 & $[0.8100,0.9521]$ & $[1.0421,1.3847]$ \\
0.005 & 0.5148 & 8.11 & 1.0627 & 1.0751 & $[0.8097,0.9521]$ & $[1.0110,1.3495]$ \\
0.01 & 0.5199 & 6.51 & 1.0618 & 1.0641 & $[0.8094,0.9521]$ & $[0.9838,1.3187]$ \\
0.02 & 0.5301 & 4.84 & 1.0596 & 1.0430 & $[0.8086,0.9519]$ & $[0.9369,1.2653]$ \\
0.03 & 0.5403 & 3.97 & 1.0568 & 1.0222 & $[0.8074,0.9515]$ & $[0.8965,1.2190]$ \\
0.05 & 0.5607 & 3.05 & 1.0479 & 0.9790 & $[0.8042,0.9498]$ & $[0.8276,1.1383]$ \\
0.2 & 0.7136 & 1.58 & 0.7672 & 0.5538 & $[0.6917,0.8190]$ & $[0.4576,0.6466]$

\end{tabular}\end{ruledtabular}
\end{table}

Table~\ref{tab:SMthermal} shows the weakening radial enhancement. For the fixed principal pair, the entropy minimum reaches unity at $n_{\rm th}=0.00695$; it is $1.0110$ at $n_{\rm th}=0.005$ and $0.9838$ at $0.01$. Thus the fixed pair loses its opposite lasting orders before $n_{\rm th}=0.01$, although both diagnostics still agree on the initial order there.

When $n_B$ is adjustable, let $n_B^{\rm tr}(n_{\rm th})$ solve $\mathcal R_{\rm tr}^+=1$ and $n_B^{\mathscr S}(n_{\rm th})$ solve $\mathcal R_{\mathscr S}^-=1$. The connected opposite-order range is $n_B^{\rm tr}<n_B<n_B^{\mathscr S}$, with both initial ratios exceeding unity. Its boundaries meet at $n_{\rm th}=0.02917$ and $n_B=0.3130$ [Fig.~\ref{fig:three}(c)]. A scan of $0\leq n_B\leq1$ with step $0.005$ at $n_{\rm th}=0$, $0.005$, $0.01$, $0.02$, $0.03$, $0.04$, and $0.05$ finds no additional component; the displayed boundaries use refined roots of the phase extrema. Continuous optimization of the margin~\eqref{eq:margin} over $n_B$ gives $4.62\%$, $3.67\%$, $2.81\%$, and $1.28\%$ at $n_{\rm th}=0$, $0.005$, $0.01$, and $0.02$, at $n_B=0.3293$, $0.3258$, $0.3226$, and $0.3172$, respectively; the two limiting margins coincide at each optimum.

Trace-interval endpoints change by less than $3.2\times10^{-13}$ between $N=24$, $48$, and $64$ in the tabulated cases. Independent full-Liouvillian checks at $N=40$ and $n_{\rm th}=0.01$, $0.05$, $0.2$ give stationary trace errors below $1.4\times10^{-14}$, and the three leading real parts agree within $6.6\times10^{-14}\omega$. These calculations set $\gamma_n=K=\gamma_2=0$; the dephasing scan of Appendix~\ref{app:finite} uses $n_{\rm th}=0$.

\section{First-order branch selection}\label{app:selection}
\subsection{Spectral projection and trace identity}
At fixed elliptic parameters, the leading even eigenvalues of $\cL_0$ are distinct and separated from the faster even spectrum. For the biorthonormal eigenoperators $\varrho_j,Y_j$ of this triple, define $\mathcal P_{\rm lead}O=\sum_j\varrho_j\Tr(Y_j^\dagger O)$ and $M$ by Eq.~\eqref{eq:M}. Regular first-order perturbation theory gives $\lambda_j^{(1)}=M_{jj}$, so $\lambda_r^{(1)}+\lambda_+^{(1)}+\lambda_-^{(1)}=\Sigma_{\mathcal V}$, the trace of $\mathcal P_{\rm lead}\mathcal V\mathcal P_{\rm lead}$ on the unperturbed spectral subspace. With $\lambda_-^{(1)}=\lambda_+^{(1)*}$ and $\lambda_r^{(1)}$ real,
\begin{equation}
 \lambda_r-\operatorname{Re}\lambda_+=\lambda_r^{(1)}-\operatorname{Re}\lambda_+^{(1)}+\ord(\epsilon^2/\omega)
 =\frac{3\lambda_r^{(1)}-\Sigma_{\mathcal V}}2+\ord(\epsilon^2/\omega).\label{eq:SMrule}
\end{equation}
Coupling to other modes contributes at higher order. The expansion follows the isolated triple at fixed $\nu>0$; Sec.~\ref{app:scope} compares these branches with the rest of the nonlinear spectrum.

The covariance coordinates $\bm v=(v_n,v_{m'},v_{m''})^T$, with $v_n=n-n_{\rm ss}$, $v_{m'}=\operatorname{Re}(m-m_{\rm ss})$, and $v_{m''}=\operatorname{Im}(m-m_{\rm ss})$, give
\begin{equation}
 M_0=\begin{pmatrix}-\kappa&0&-4\xi\\0&-\kappa&2\omega\\-4\xi&-2\omega&-\kappa\end{pmatrix}.
\end{equation}
For its real branch, $v_{m''}=0$ and $v_{m'}=-2\xi v_n/\omega$; the left eigenoperator is $H_\xi$, with energy deviation $v_E=\omega v_n+2\xi v_{m'}=(\nu^2/\omega)v_n$. The right eigenoperators are derivatives of $G[W]$. Differentiated Gaussian moment identities therefore compute matrix elements of polynomial perturbations on this subspace, including perturbations whose full dynamics is non-Gaussian.

\subsection{Dephasing, Kerr interaction, and loss}
Number dephasing adds $\gamma_n\operatorname{diag}(0,-2,-2)$ to the exact second-moment matrix, giving
\begin{equation}
 \lambda_r^{(1)}=\frac{8\xi^2}{\nu^2}\gamma_n,\qquad \Sigma_{\mathcal V}=-4\gamma_n,
 \qquad \delta=2\left(1+\frac{6\xi^2}{\nu^2}\right)\gamma_n+\ord(\gamma_n^2/\omega).
\end{equation}
The exact characteristic equation, with $\ell=\lambda+\kappa$, is
$\ell[(\ell+2\gamma_n)^2+4\omega^2]-16\xi^2(\ell+2\gamma_n)=0$. Its real root has an exact left eigenoperator quadratic in $a$ and $a^\dagger$, which tends to $H_\xi/\omega$ as $\gamma_n\downarrow0$; this explains the energy ratio selected in this limit.

For the Kerr interaction, the moment equations obey $\dot n|_K=0$ and $\dot m|_K=-iK\langle 2a^\dagger a^3+a^2\rangle$. Differentiating $\langle a^\dagger a^3\rangle=3nm$ at the unperturbed stationary Gaussian gives the projected matrix per unit $K$,
\begin{equation}
 M_K=\begin{pmatrix}0&0&0\\6m''_{\rm ss}&0&6n_{\rm ss}+1\\-6m'_{\rm ss}&-(6n_{\rm ss}+1)&0\end{pmatrix},
\end{equation}
and the energy pairing gives
\begin{equation}
 \lambda_r^{(1)}=-\frac{24\kappa\omega\xi^2}{\nu^2\Gamma}K=-\alpha K,
 \qquad \Sigma_{\mathcal V}=K\Tr M_K=0,
 \qquad \delta=-\tfrac32\alpha K+\ord(K^2/\omega).
\end{equation}
For nonzero $\xi$, ${\rm sgn}\,\alpha={\rm sgn}\,\omega$. The full Liouvillian at $N=28$ confirms the exact symmetry of Sec.~\ref{sec:model}: at $(\omega,K)=(1,10^{-3})$ and $(-1,-10^{-3})$ it gives the same $\lambda_r=-0.500030$ and $\operatorname{Re}\lambda_+=-0.499985$, and the other two sign combinations both give $\lambda_r=-0.499969$ and $\operatorname{Re}\lambda_+=-0.500016$.

Two-photon loss is treated with the same method. The adjoint identities $\cD[a^2]^\dagger n=-2a^{\dagger2}a^2$ and $\cD[a^2]^\dagger a^2=-(2n+1)a^2$, together with $\langle a^{\dagger2}a^2\rangle=2n^2+|m|^2$ and $\langle a^\dagger a^3\rangle=3nm$, give
\begin{equation}
 M_2=\begin{pmatrix}
 -8n_{\rm ss}&-4m'_{\rm ss}&-4m''_{\rm ss}\\
 -6m'_{\rm ss}&-(6n_{\rm ss}+1)&0\\
 -6m''_{\rm ss}&0&-(6n_{\rm ss}+1)
 \end{pmatrix}.
\end{equation}
For $\mathcal V=\gamma_2\cD[a^2]$,
\begin{equation}
 \lambda_r^{(1)}=\gamma_2\left(\frac{4\xi^2}{\nu^2}-\frac{48\xi^2}{\Gamma}\right),\qquad
 \Sigma_{\mathcal V}=-\gamma_2(20n_{\rm ss}+2),
\end{equation}
which gives Eq.~\eqref{eq:loss2}. Balanced thermal noise $\gamma_T(\cD[a]+\cD[a^\dagger])$ changes only diffusion, giving $\lambda_r^{(1)}=\Sigma_{\mathcal V}=0$. A heating term $\gamma_h\cD[a^\dagger]$ alone changes the net damping to $\kappa-\gamma_h$, so $\lambda_r^{(1)}=\gamma_h$, $\Sigma_{\mathcal V}=3\gamma_h$, and again $\delta=0$. The projected matrix elements for all examples agree with direct left/right Liouvillian overlaps to within $2.5\times10^{-15}$ at strength $10^{-3}$.

\subsection{Selected ratios and the separation scale}
For a selected complex pair, both conjugate eigenoperators contribute. If $\varrho$ denotes one member and $c_i$ its preparation coefficient, the ratio is
\begin{equation}
 \mathcal R_{\rm pair}(\vartheta)=\frac{\|c_Ae^{i\vartheta}\varrho+c_A^*e^{-i\vartheta}\varrho^\dagger\|_1}
 {\|c_Be^{i\vartheta}\varrho+c_B^*e^{-i\vartheta}\varrho^\dagger\|_1}.
\end{equation}
The selected ratios quoted in Sec.~\ref{sec:selection} use $N=44$. For simultaneous number dephasing and Kerr interaction,
\begin{equation}
 \delta=2(1+6\xi^2/\nu^2)\gamma_n-\tfrac32\alpha K
 +\ord((K^2+|K|\gamma_n+\gamma_n^2)/\omega),
\end{equation}
whose first-order zero is $K/\omega=46.4\gamma_n/\omega$ for the main parameters. Suppressing the subdominant contribution below a relative tolerance $\eta$ takes a time of order $|\delta|^{-1}\log(1/\eta)$, up to an overlap-dependent constant. At strength $10^{-3}\omega$, $\omega/|\delta|$ is approximately $471$ for dephasing and $22\,070$ for same-sign Kerr. A displacement of amplitude $\beta$ introduces an odd contribution of order $|\beta|e^{-\kappa t/2}$, overtaking an even tail of amplitude $C_{\rm e}e^{-\kappa t}$ around $t\simeq2\log(C_{\rm e}/|\beta|)/\kappa$.

\subsection{Accuracy and nonlinear spectral scope}\label{app:scope}
Figure~\ref{fig:two}(d) compares the full splitting between the continued covariance branches with the first-order rule at $N=34$. For the Kerr interaction the ratio $\delta/\delta^{(1)}$ is $0.9884$ and $1.0118$ at $|K|=10^{-3}\omega$ for the two signs; for dephasing it differs from unity by less than $7\times10^{-4}$ up to $\gamma_n=0.1\omega$.

The semiclassical amplitude obeys
$\dot z=-[\kappa/2+i(\omega+K|z|^2)]z-2i\xi z^*$, so a nonzero stationary amplitude with $n=|z|^2$ requires
\begin{equation}
 (\omega+Kn)^2+(\kappa/2)^2=4\xi^2.\label{eq:SMnonlinear}
\end{equation}
The main parameters satisfy $\kappa>4|\xi|$, so this condition has no solution. The exact moment inequality
$d\langle n\rangle/dt\leq-(\kappa-4|\xi|)\langle n\rangle+2|\xi|$,
which follows from $|\langle a^2\rangle|\leq\sqrt{\langle n\rangle(\langle n\rangle+1)}$, bounds excitation growth independently of $K$. Sparse spectral checks at $N=48$, $64$, and $96$ retain the covariance real branch as the leading even decay for $K=-0.001\omega$ and $-0.03\omega$. At the latter strength its eigenvalue is $-0.49897\omega$, while an additional real branch converges near $-0.59798\omega$.

For the second generator, $\xi=0.15\omega$ and $\kappa=0.5\omega$, opposite-sign Kerr admits semiclassical stationary occupations $(1\pm\sqrt{0.0275})/|K|$ (in units with $\omega=1$). At $K=-0.03\omega$, a slow even Liouvillian eigenvalue changes from $-0.17761\omega$ at $N=40$ to $-0.0072523\omega$ at $N=96$ and $128$, whereas the continued covariance real branch is near $-0.49734\omega$ [Fig.~\ref{fig:SMchecks}(c)]. At smaller $|K|$, the stationary occupations move to higher photon number, so resolving the slow branch requires larger Fock cutoffs.

\section{Finite-time dephasing and measurement accuracy}\label{app:finite}
Table~\ref{tab:SMtolerance} gives full density-operator results for $\cL_0+\gamma_n\cD[n]$. The first crossing and first return bound the continuous reversal window; further crossings may precede the last crossing into the real-branch order. The scan ends at $\omega t=80$, giving lower bounds for $\gamma_n/\omega=0.001$ and $0.003$. The last-crossing times $|\delta|t$ are $0.625$ at $\gamma_n=0.05\omega$ and $0.798$ at $0.01\omega$.
\begin{table}[ht]
\caption{Full-Liouvillian trace-distance crossings for $n_B=0.33$, $N=34$, and $0\leq\omega t\leq80$, in units of $1/\omega$. The last crossing is followed by the selected real-branch order; ``None'' at $\gamma_n=0$ means lasting reversal. At $\gamma_n=0.1\omega$ no crossing occurs, so A remains farther throughout; dashes mark inapplicable entries. Distances below $10^{-15}$ are obtained with the shifted propagation of Appendix~\ref{app:numerics}.}\label{tab:SMtolerance}
\begin{ruledtabular}\begin{tabular}{lllll}
$\gamma_n/\omega$ & First crossing & First return & Last crossing & $D_B(t_f)$\\\hline
0 & 1.358 & None & None & --- \\
0.001 & 1.364 & $>80$ & $>80$ & --- \\
0.003 & 1.376 & 29.405 & $>80$ & --- \\
0.005 & 1.389 & 16.680 & 73.011 & $6.2\times10^{-17}$ \\
0.01 & 1.420 & 3.583 & 37.577 & $3.1\times10^{-9}$ \\
0.02 & 1.483 & 3.339 & 18.469 & $4.3\times10^{-5}$ \\
0.03 & 1.549 & 3.175 & 12.103 & $1.0\times10^{-3}$ \\
0.05 & 1.691 & 2.907 & 5.885 & $2.2\times10^{-2}$ \\
0.1 & No crossing & --- & --- & ---

\end{tabular}\end{ruledtabular}
\end{table}

For the unperturbed pair, the trace crossing is at $\omega t=1.3578$. Relative entropy crosses at $1.8900$ and $2.6190$ and places A farther thereafter, consistent with its strict leading interval. HS crosses repeatedly, including at $3.1858$ and $4.4476$. The refined near-touching extrema are $R_{\rm tr}=0.98388$ at $\omega t=3.8193$ and $R_{\mathscr S}=1.00464$ at $5.4609$. At the last entropy crossing, $(D_A,D_B)=(0.10569,0.11748)$. At $\omega t=3$, $(D_A,D_B)=(0.090769,0.098027)$ and $(\mathscr S_A,\mathscr S_B)=(0.159103,0.148538)$, while $D_{{\rm HS},A}-D_{{\rm HS},B}=-2.36\times10^{-3}$. At $\omega t=4$, the differences are $D_A-D_B=-1.23\times10^{-3}$, $\mathscr S_A-\mathscr S_B=+1.47\times10^{-2}$, and $D_{{\rm HS},A}-D_{{\rm HS},B}=+3.12\times10^{-3}$.

If reconstructed states have trace-distance errors $e_A,e_B$ and their common reference has error $e_s$, the triangle inequality bounds the uncertainty in $D_A-D_B$ by $e_A+e_B+2e_s$. If each reported distance already includes its reference uncertainty and has total error at most $e_D$, the bound becomes $2e_D$, and a measured sign is certified when $|D_A-D_B|>2e_D$. Relative entropy requires its own propagated error $e_{\mathscr S}$, since its sensitivity depends on the stationary spectrum.

For Gaussian reconstruction, let each independent element of $V_A(t)$, $V_B(t)$, and $W$ have error at most $e_V$. The largest linearized error in $X_A-X_B$ is $e_V\sigma_X$, where $\sigma_X$ sums the absolute derivatives with respect to $V_A$, $V_B$, and the common $W$. At $\omega t=3$, $\sigma_{\mathscr S}=6.2522$ and $\sigma_{\rm tr}=2.5919$. Dividing the signals in Eq.~\eqref{eq:signal} by these sensitivities gives local precision scales $1.6898\times10^{-3}$ and $2.8000\times10^{-3}$. The entropy derivatives with respect to $W$ sum to $10.6$ and $10.0$ for A and B, but their difference contributes only $1.3$. Using a common reference therefore suppresses much of this uncertainty. Trace distances use Fock cutoff $N=70$, entropies use Eq.~\eqref{eq:SMREexact}, and derivatives are central differences.

\section{Numerical procedures and convergence}\label{app:numerics}
Hermitian trace norms use all eigenvalues in both photon-number parity blocks. Liouvillians use column-major vectorization and the full even operator sector. Gaussian tails are evaluated before multiplying by their decay factor, and periodic extrema are located on a phase grid and refined, including extrema that cross the phase seam. To propagate very small deviations, we eliminate the vacuum diagonal entry by the exact trace-zero constraint, shift the restricted generator by its leading real part, and use matrix-exponential action, multiplying the decay factor back only when an absolute distance is needed. Finite-time dephasing trajectories use a time step of $0.05/\omega$; cubic interpolation locates sign changes, with a doubled-step comparison and an independent cutoff check.

The principal trace interval is unchanged to eight digits for $N=20$--$36$, and the same-sign Kerr interval changes by at most $10^{-8}$ for $N\geq28$. Trace classification boundaries agree between $N=20$, $32$, and $48$ to below $10^{-12}$. Gaussian covariance evolution and independently propagated full Lindblad trace distances agree to $1.6\times10^{-15}$, and the HS coefficient from the operator tangent and from the Gaussian closed form agree to $3.4\times10^{-16}$. For the $\gamma_n=0.05\omega$ last-crossing example, increasing $N$ from $34$ to $42$ and halving the step changes the four crossing times by at most $8.8\times10^{-8}$. The full crossing checks, Liouvillian residuals, larger-cutoff spectra, and convergence data are available in the Zenodo repository~\cite{He2026Data}.

\twocolumngrid
\bibliography{references}
\end{document}